\documentclass[aps,12pt,final,notitlepage,oneside,onecolumn,nobibnotes,nofootinbib,
superscriptaddress,noshowpacs,centertags,showpacs,showkeys]{revtex4}
\usepackage{bm}
\usepackage{graphics}
\usepackage{rotating}
\usepackage{epsfig}
\usepackage[english,russian]{babel}

\begin{document}
\selectlanguage{english}

\title{Resonant states of muonic three-particle systems with lithium, 
helium and hydrogen nuclei}
\author{\firstname{A. V.} \surname{Eskin}}
\affiliation{Samara University, Moskovskoye Shosse 34, 443086, Samara, Russia}
\author{\firstname{V. I.} \surname{Korobov}}
\affiliation{BLTP JINR, Joliot-Curie 6, 141980, Dubna, Moscow Region, Russia}
\affiliation{Samara University, Moskovskoye Shosse 34, 443086, Samara, Russia}
\author{\firstname{A. P.} \surname{Martynenko}}
\affiliation{Samara University, Moskovskoye Shosse 34, 443086, Samara, Russia}
\author{\firstname{F. A.} \surname{Martynenko}}
\affiliation{Samara University, Moskovskoye Shosse 34, 443086, Samara, Russia}

\begin{abstract}
We study the energy spectrum of three-particle systems $(He-p-\mu)$, 
$(He-d-\mu)$, $(Li-p-\mu)$ and $(Li-d-\mu)$ on the basis of variational approach 
with exponential 
and Gaussian basis. Using the Complex Coordinate Rotation (CCR) method we
calculate energies of resonant states of listed molecules.
\end{abstract}



\maketitle

\section{Introduction}
\label{intro}

In this paper we consider three-particle muonic systems of the type $(\mu p He)$, 
$(\mu d He)$, $(\mu p Li)$, $(\mu d Li)$ which 
include nuclei of hydrogen isotopes, as well as helium and lithium nuclei. 
The study of the reactions of formation of such molecules and their characteristics 
is important for calculating the probabilities of muon catalysis reactions.
As shown by previous studies, these particles can form molecular three-particle resonances
\cite{kravtsov,belyaev,kravtsov1,bystritskii,korobov}. 
Resonances $(d-He-\mu)$ were observed in collisions of muonic hydrogen atoms with helium nuclei. 
In the case of nuclei with a large charge, there is a stronger repulsion (proportional 
to the charge of the heavy nucleus), which, however, is partially compensated by the 
attractive force due to the polarizability of hydrogen atoms. 
Various methods are used to calculate the energies of three-particle systems, 
including the method of hyperspherical functions.
In what follows we will call the states $(p-He-\mu)$, $(d-He-\mu)$, $(p-Li-\mu)$, 
$(d-Li-\mu)$ either bound or resonant states, although they can be called resonant 
states since they are embedded in the continuum $(\mu He)+p$, $(\mu He)+d$, etc.
In this paper, we propose 
to use the method of complex coordinate rotation \cite{ho,WR} for calculating resonance states, 
which has been successfully used to solve many problems in the physics of three-particle 
systems. To solve the problem by the variational method, two bases of exponential 
and Gaussian functions are used \cite{var1,var2,var3,var4}.
The aim of the work is to calculate the energy levels of the ground states in 
the listed three-particle systems using different basis functions in the 
non-relativistic approximation and to calculate the main corrections to the 
non-relativistic result. 
Precision research into the energy levels of exotic atoms and molecules is 
a research area in theoretical physics that allows, among other things, 
to determine more precise values of fundamental parameters.

\section{General formalism}
\label{formalism}

To calculate the energy levels of three-particle molecules, we introduce 
the Hamiltonian of the system, which in the center-of-mass system in the 
non-relativistic approximation is constructed from paired Coulomb interactions 
of particles:
\begin{equation}
\label{f1}
H = T+V =
\frac{\mathbf{p}_1^2}{2m_1}+\frac{\mathbf{p}_2^2}{2m_2}+\frac{\mathbf{p}_3^2}{2m_3}
+\frac{z_1z_2}{R}+\frac{z_1z_3}{r_1}+\frac{z_2z_3}{r_2},
\end{equation}
where $T$ and $V$ denote the kinetic and potential energy operators, 
and the particle momenta satisfy
the relation in the center-of-mass reference frame: 
$\mathbf{p}_1\!+\!\mathbf{p}_2\!+\!\mathbf {p}_3=0$. 
The particles are numbered as follows:
1 - for the helium or lithium nucleus, 2 - for the muon and 3 - for the proton or
deuteron, correspondingly. The particle charges are denoted as $z_a$ ($a=1,2,3$), 
and $\mathbf{R}=\mathbf{r}_{12}=\mathbf{R}_2-\mathbf{R }_1$,
$\mathbf{r}_1=\mathbf{r}_{13}=\mathbf{R}_3-\mathbf{R}_1$, $\mathbf{r}_2=\mathbf{r} _{23}=\mathbf{R}_3-\mathbf{R}_2$,
where $\mathbf{R}_a$ are the coordinates of the particles in the center-of-mass system.

To determine the energy levels in this case, the complex coordinate rotation (CCR) 
method can be used~\cite{ho}. Using this method,
the following complex transformation of the Hamiltonian is performed:
\begin{equation}
\label{f2}
H\rightarrow H(\theta)=T\exp(-2i\theta)+V\exp(-i\theta)\,,
\end{equation}
which arises under the corresponding coordinate transformation: 
$r\rightarrow r\exp(i\theta)$,
where the rotation angle $\theta$ is real and positive. In the complex energy plane, 
for sufficiently large values of $\theta$, this transformation rotates the continuous 
spectrum of the Hamiltonian, exposing resonant poles around the thresholds 
from the unphysical sheet to the physical sheet of the Riemann surface, while the bound 
state poles remain unchanged in the negative part of the real energy axis. 
It was shown that resonances arise as complex energy levels and the corresponding 
wave functions are square integrable.

To calculate the energy levels using the variational method, two sets of trial wave 
functions were used: exponential and Gaussian. Thus, in the case of the Gaussian basis, 
the trial wave function has the form \cite{apm2024,apm2024a,apm2023}:
\begin{equation}
\label{f3}
\Psi({\boldsymbol\rho},{\boldsymbol\lambda},A^i_{ij})=\sum_{i=1}^K C_i 
\psi_i({\boldsymbol\rho},{\boldsymbol\lambda},A_{ij})=
\sum_{i=1}^K C_i e^{-\frac{1}{2}[A^i_{11}{\boldsymbol\rho}^2+2A^i_{12}{\boldsymbol\rho}
{\boldsymbol\lambda}+A_{22}^i
{\boldsymbol\lambda}^2]},
\end{equation}
where the $C_i$ variables are linear variational parameters.
The Jacobi coordinates ${\boldsymbol\rho}$ and ${\boldsymbol\lambda}$ are related 
to the radius vectors ${\bf r}_1$, ${\bf r}_2$, and ${\bf r}_3$ as follows:
\begin{equation}
\label{f4}
{\boldsymbol\rho}={\bf r}_2-{\bf r}_1,~~~
{\boldsymbol\lambda}={\bf r}_3-\frac{m_1{\bf r}_1+m_2{\bf r}_2}{m_1+m_2},
\end{equation}
where $A_{ij}$ is the matrix of nonlinear parameters. The problem is to find such values 
of the parameters and expansion coefficients that the average value of the Hamiltonian 
was minimal.
To find the energies of bound states, the Schr\"odinger equation with the Coulomb 
interaction of three 
particles is reduced to solving a matrix eigenvalue problem of the form:
\begin{equation}
\label{f5}
HC=E^\lambda BC,
\end{equation}
where the matrix elements of the Hamiltonian $H_{ij}=<\psi_i|H|\psi_j>$ and 
the normalization $B_{ij}=<\psi_i|\psi_j>$ 
are calculated analytically using the variational wave functions, and 
$E^\lambda$ is one of the energy eigenvalues.
The upper bound for the state energy of a system of three particles in the variational approach was provided by the smallest eigenvalue of the generalized eigenvalue problem.

The matrix elements of the normalization have the following form:
\begin{equation}
\label{eq9}
\langle \psi'|\psi\rangle =\mathrm{\int\!\!\!\!\int}
{d\boldsymbol{\mathrm{\rho }}}d\boldsymbol{\mathrm{\lambda }}e^{-\frac{1}{2}
\left[B_{11}{\boldsymbol{\mathrm{\rho }}}^{\mathrm{2}}+
B_{22}{\boldsymbol{\mathrm{\lambda }}}^{\mathrm{2}}+2B_{12}\left(\boldsymbol{\mathrm{\rho }}
\boldsymbol{\mathrm{\lambda }}\right)\right]},
\end{equation} 
where the matrix of coefficients $B = A + A'$. Let us turn to spherical coordinates and denote 
the angle between vectors \textbf{$\boldsymbol{\rho}$} and \textbf{$\boldsymbol{\lambda}$} 
as \textbf{$\boldsymbol{\theta }$}. Then, integrating over particle coordinates, we obtain:
\begin{equation}
\label{eq10}
\left\langle \psi'|\psi\right\rangle =
8{\pi }^2\int^{\infty }_0{\int^{\infty }_0{\int^{\pi }_0{{\mathrm{\rho }}^{\mathrm{2}}
{\mathrm{\lambda }}^{\mathrm{2}}}}}d\mathrm{\rho }d\mathrm{\lambda }
d\left(cos\mathrm{\theta }\right)e^{-\frac{1}{2}\left[B_{11}{\mathrm{\rho }}^{\mathrm{2}}+B_{22}{\mathrm{\lambda }}^{\mathrm{2}}+2B_{12}\mathrm{\rho }\mathrm{\lambda }\mathrm{cos}\mathrm{\theta }\right]}=\frac{8{\pi }^3}{{{\mathrm{(}\mathrm{det} B)\ }}^{\frac{3}{2}}}
\end{equation} 

To calculate the matrix elements of the Hamiltonian, we write out expanded expressions for 
the kinetic and potential energies of the system in Jacobi coordinates.
Then the matrix elements of kinetic energy have the following analytical form:
\begin{equation}
\label{eq13}
\left\langle \psi'\left|\hat{T}
\right|\psi\right\rangle =
-\frac{24{\pi }^3}{{{\mathrm{(}\mathrm{det} B)\ }}^{{5}/{2}}}
\left[\frac{{\hbar}^2}{2{\mu }_1}I_{\rho }+\frac{{\hbar}^2}
{2{\mu }_2}I_{\lambda }\right],
\end{equation} 
\begin{equation}
\label{eq14}
I_{\rho }=A^2_{12}B_{11}-2A_{11}A_{12}B_{12}+A_{11}\left(B^2_{12}+
\left(A_{11}-B_{11}\right)B_{22}\right),
\end{equation} 
\begin{equation}
\label{eq15}
I_{\lambda }=A^2_{12}B_{22}-2A_{22}A_{12}B_{12}+A_{22}\left(B^2_{12}+
\left(A_{22}-B_{22}\right)B_{11}\right),
\end{equation} 
where the reduced masses are presented as ${\mu }_1=\frac{m_1m_2}{m_1{+m}_2}$, 
${\mu }_2=\frac{(m_1+m_2)m_3}{m_1{+m}_2{+m}_3}$. 

The matrix elements of potential energy have the following form ($m_{12}=m_1+m_2$):
\begin{equation}
\label{eq22}
\left\langle\psi'\left|\hat{V}\right|\psi\right\rangle =
z_1z_2I_{12}+z_1z_3I_{13}+z_2z_3I_{23}, 
\end{equation} 
\begin{equation}
\label{eq23}
I_{12}=\frac{8\sqrt{2}{\pi }^{{5}/{2}}}{\sqrt{B_{22}}
{\mathrm{det} B\ }}\ ,\ I_{13}=\frac{8\sqrt{2}{\pi }^{{5}/{2}}}{\sqrt{F^{13}_1}
\left(B_{22}F^{13}_1-{\left(F^{13}_2\right)}^2\right)},
I_{23}=\frac{8\sqrt{2}{\pi }^{{5}/{2}}}{\sqrt{F^{23}_1}\left(B_{22}F^{23}_1-
{\left(F^{23}_2\right)}^2\right)}.
\end{equation} 

For numerical calculation, a program is written in the MATLAB system for solving 
the three-particle Coulomb problem in the framework of the stochastic variational method
on the basis of Fortran program from \cite{var1}.

To improve the accuracy of the energy level calculations, we take into account 
the contribution of a number of other terms in the spin-independent part of 
the Breit-Pauli Hamiltonian \cite{BS,BLP}, which have the following form:
\begin{eqnarray}
\label{f6}
H^{(2)} = \alpha^2\Bigg\{
-\frac{\mathbf{p}_1^4 }{8m_1^3} - \frac{\mathbf{p}_2^4}{8m_2^3} - \frac{\mathbf{p}_3^4}{8m_3^3}
 + \frac{z_1 z_2\pi}{2m_\mu^2}  
\delta(\mathbf{r}_2-\mathbf{r}_1)+
\frac{z_2 z_3\pi}{2m_\mu^2}  \delta(\mathbf{r}_3-\mathbf{r}_2)-
\\
-\sum_{a>b}\frac{z_az_b}{2m_am_b}\left[\frac{\mathbf{p}_a\cdot\mathbf{p}_b}{r_{ab}}
+\frac{\mathbf{r}_{ab}\cdot(\mathbf{\bf r}_{ab}\cdot\mathbf{p}_a)\mathbf{p}_b}{r_{ab}^3}\right]
\Bigg\}.  \nonumber  
\end{eqnarray}
The contribution of these terms of the interaction operator to the energy spectrum 
is calculated using the transformed wave function within the CCR method.

Our calculations also take into account the correction for the nuclear finite size, 
which is determined by the matrix elements of the $\delta$-functions in the form 
\cite{SapYen90}:
\begin{equation}
\label{f7}
E^{(2)}_{\rm nuc}= \frac{2\pi z_1\left(\frac{R_{\rm He,Li}}{a_0}\right)^2}{3}
\langle\delta(\mathbf{r}_1)\rangle\,,
\end{equation}
where $R_{\rm He, Li}$ is the root-mean-square (rms) radius of the charge distribution 
of the nucleus $He$ or $Li$. The total contribution of \eqref{f6} 
and \eqref{f7} is denoted below by $E^{(2)}$:
\begin{equation}
\label{f8}
E^{(2)} = \langle H^{(2)}\rangle + E^{(2)}_{\rm nuc}\,.
\end{equation}

The results of the calculation of non-relativistic energies and corrections 
to them are presented in two Tables~\ref{tb1},\ref{tb2}.
The first two lines contain the energies of three-particle systems obtained in the 
nonrelativistic approximation using exponential and Gaussian trial functions. 
The remaining lines present relativistic corrections, corrections for recoil, 
nuclear structure, and contact interaction, which are obtained using exponential 
basis functions.

\begin{table}[htbp]
\caption{Numerical values of bound state energies and leading order corrections 
for molecules $(He-\mu-p)$, $(Li-\mu-p)$}
\bigskip
\begin{center}
\begin{tabular}{|c|c|c|c|c|} \hline 
 & ${}^3{He}p\mu $ & ${}^4{He}p\mu $ & ${}^6{Li}p\mu $ & ${}^7{Li}p\mu $ \\ \hline 
Energy(Exp. Basis) & -95.6365412927 & -95.9279137052 & -93.5991296303 &-93.6340512645 \\ \hline
Energy(Gaussian Basis) &-95.6628556322 &-95.9426289157 &-93.5988425958 & -93.6354350556 \\ \hline 
Rel. corr. & $-0.0053592295$ & $-0.0054660024$ & $-0.0049689089$ & $-0.0049911010$ \\ \hline 
Recoil corr. & $0.0009275814$ & $0.0008792477$ & $0.0008676392$ & $0.0008616638$ \\ \hline 
Structure corr. & $0.0000000554$ & $0.0000000476$ & $0.0000000378$ & $0.0000000433$ \\ \hline 
Contact corr. & $0.0000000005$ & $0.0000000005$ & $0.0000000004$ & $0.0000000004$ \\ \hline 
\end{tabular}
\end{center}
\label{tb1}
\end{table}
\begin{table}[htbp]
\caption{Numerical values of bound state energies and leading order corrections 
for molecules $(He-\mu-d)$, $(Li-\mu-d)$}
\bigskip
\begin{center}
\begin{tabular}{|c|c|c|c|c|} \hline 
 & ${}^3{He}d\mu $ & ${}^4{He}d\mu $ & ${}^6{Li}d\mu $ & ${}^7{Li}d\mu $ \\ \hline 
Energy(Exp. Basis)&-100.4791108238& -100.7897983021& -98.6171240799 &-98.6592929873\\ \hline 
Energy(Gaussian Basis) &-100.4448590779 &-100.7440987563 &-98.6183973574 &-98.6571724625 \\ \hline 
Rel. corr. & $-0.0061926389$ & $-0.0062871348$ & $-0.0059525477$ & $-0.0049911554$ \\ \hline 
Recoil corr. & $0.0001868027$ & $0.0001534442$ & $0.0000353593$ & $0.0000324794$ \\ \hline 
Structure corr.& $0.0000001923$ & $0.0000001841$ & $0.0000001918$ & $0.0000001904$ \\ \hline 
Contact corr.& $0.0000000005$ & $0.0000000005$ & $0.0000000005$ & $0.0000000005$ \\ \hline 
\end{tabular}
\end{center}
\label{tb2}
\end{table}

\section{Results and discussion}
\label{concl}

In this paper, we have investigated the energy levels of three-particle muonic 
molecules with helium, lithium, and hydrogen nuclei. In such systems, the two 
particles have a large positive charge, which leads to a repulsive potential. 
However, as studies in previous papers have shown, there are resonant states 
in such systems. Since the results of previous calculations differ from each 
other, it was interesting to perform the calculation using a different method, 
with a different basis.

\begin{figure}[htbp]
\centering
\includegraphics[scale=0.55]{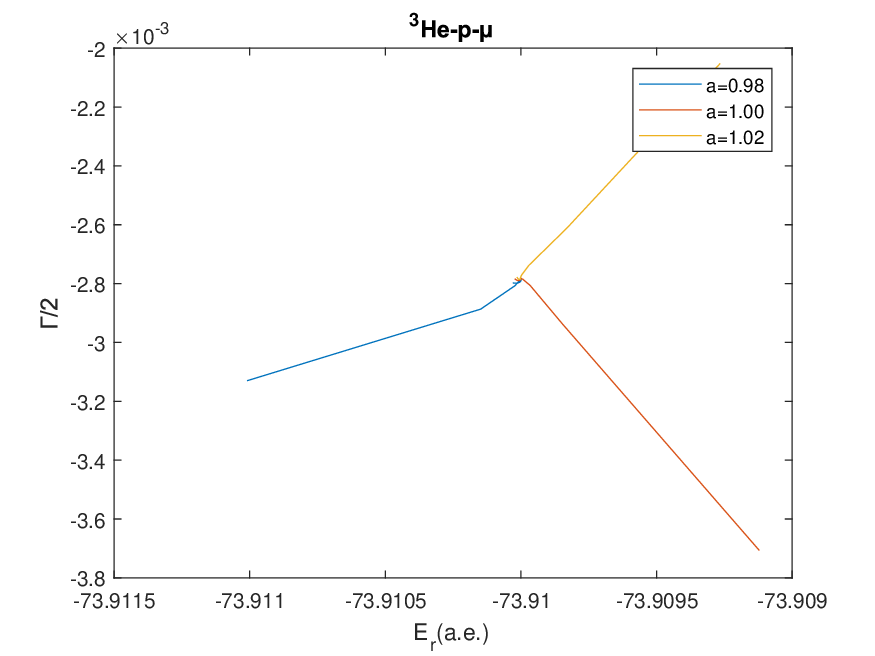}
\includegraphics[scale=0.55]{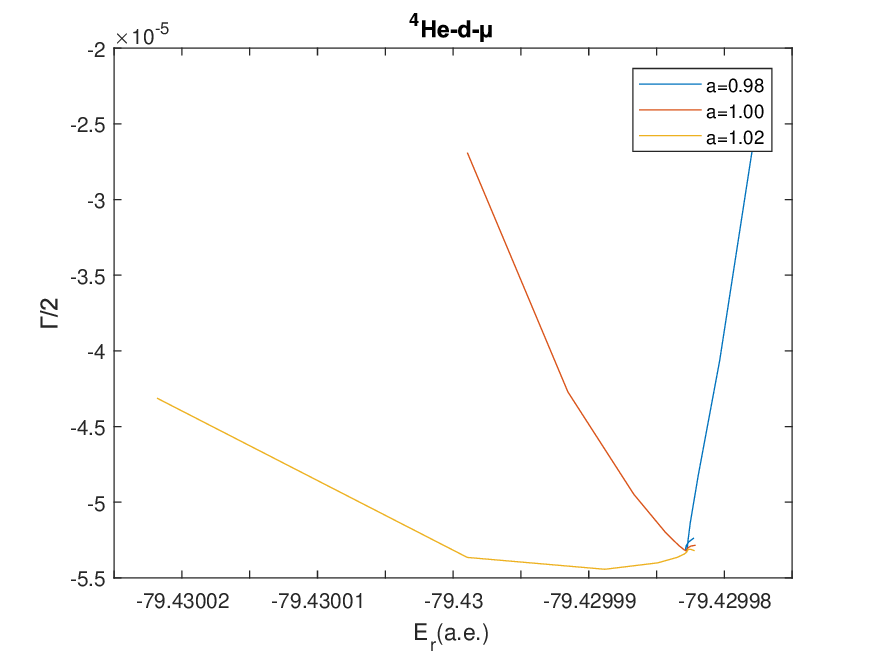}
\includegraphics[scale=0.55]{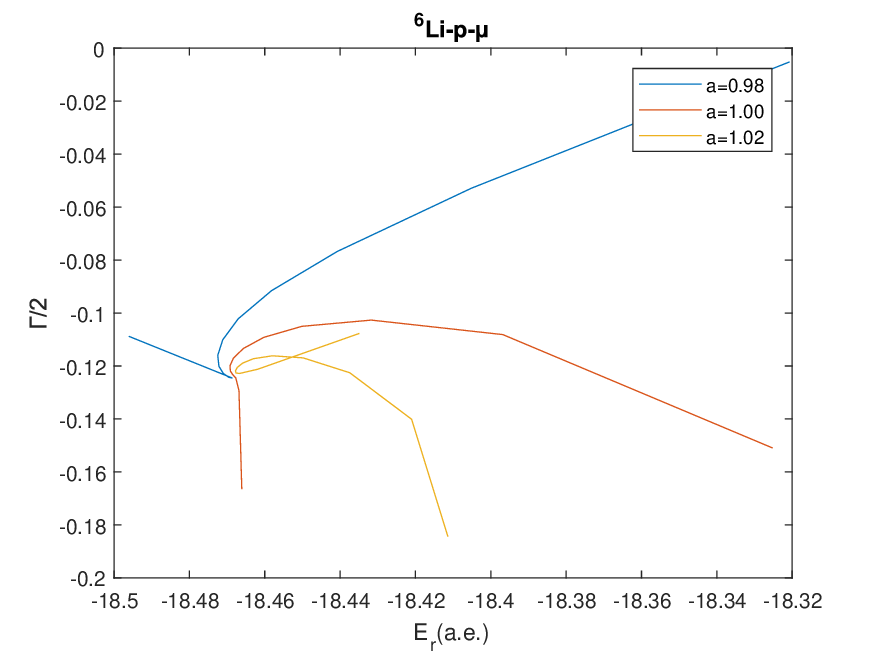}
\includegraphics[scale=0.55]{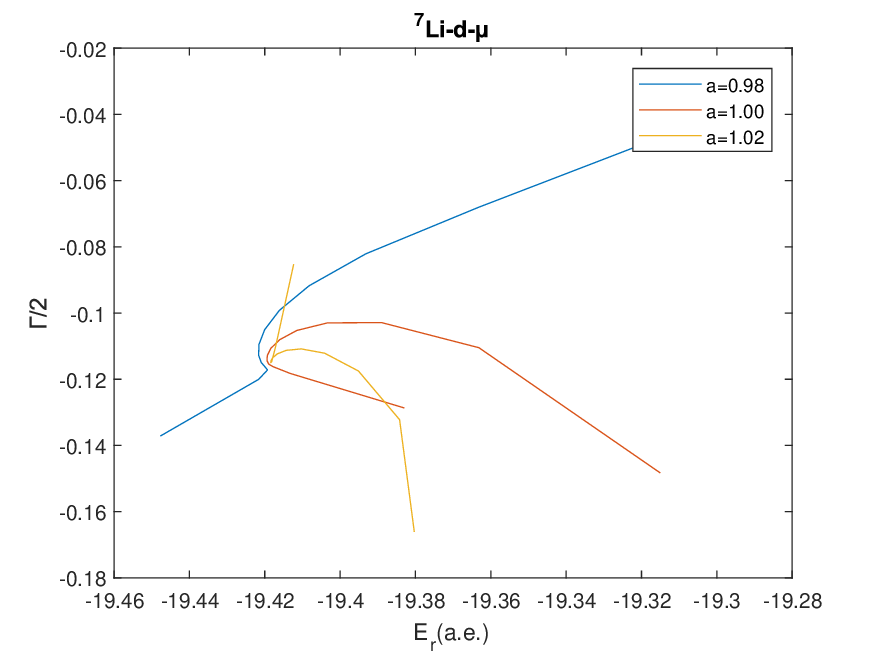}
\caption{Rotational paths for resonance in the systems
$(He-\mu-p)$, $(He-\mu-d)$, $(Li-\mu-p)$, $(Li-\mu-d)$. 
The node in the center of the graph corresponds to the stationary point defining 
the position of the resonance $E$ in the complex energy plane.}
\label{fig1}
\end{figure}

The results we obtained within the framework of the complex coordinate rotation method are presented both in Tables~\ref{tb1},\ref{tb2} and in Fig.~\ref{fig1}.
The idea of the method is to transform coordinates of the form ${\bf r}\to{\bf r} e^\theta$. 
For each value of the dilation parameter $a_{dl}=e^{Re[\theta]}=(0.98,1.00,1.02)$, the program 
for calculating the energy of the resonance state is launched for different values of the 
rotation parameter $\phi=Im\theta=(0.1,0.12,...,0.28,0.3)$. In our calculations, the size 
of the basis N=1500 is chosen. Based on the obtained results, three curves are constructed 
on a plane, along the axes of which the real part and the imaginary part of the energy, 
called the resonance half-width, are plotted. The graph shows the point to which these 
three curves tend. This point will correspond to the resonance position. Having fixed the 
values of the dilation parameter and the rotation parameter, it is possible to calculate 
various corrections.

To find the binding energy between the heaviest positively charged particle (He or Li) 
and a neutral two-particle cluster ($p\mu$ or $d\mu$), it is necessary to use the formula:
\begin{equation}
\label{9}
\varepsilon_{bind}=-(E+E_{cl}),
\end{equation}
where $E$ is the energy of three-particle state, $E_{cl}=-\frac{\mu}{2n^2}$ in electron
atomic units ($\mu$ is the reduced mass, $n$ is the principal quantum number). 
For the systems listed in the Tables~\ref{tb1},\ref{tb2}, the binding energies are equal to:
\begin{equation}
\label{10}
{\varepsilon }_{bind}\left(^3{Hep\mu }\right)=-73.762~ \mathrm{eV},~~~
{\varepsilon }_{bind}\left(^4{Hep\mu }\right)=-81.675~ \mathrm{eV},
\end{equation}
\begin{equation}
\label{11}
{\varepsilon }_{bind}\left(^6{Lip\mu }\right)=-18.432~ \mathrm{eV},~~~
{\varepsilon }_{bind}\left(^7{Lip\mu }\right)=-19.380~ \mathrm{eV},
\end{equation}
\begin{equation}
\label{12}
{\varepsilon }_{bind}\left(^3{Hed\mu }\right)=-70.834~ \mathrm{eV},~~~
{\varepsilon }_{bind}\left(^4{Hed\mu }\right)=-79.272~ \mathrm{eV},
\end{equation}
\begin{equation}
\label{13}
{\varepsilon }_{bind}\left(^6{Lid\mu }\right)=-20.268~ \mathrm{eV},~~~
{\varepsilon }_{bind}\left(^7{Lid\mu }\right)=-21.413~ \mathrm{eV}.
\end{equation}

The obtained results for the binding energy are consistent with the results from \cite{belyaev,kravtsov}. 
However, it should be noted that the difference is quite significant for this type 
of calculation and is about 10\%.
Calculations using the variational method provide high accuracy, but we write out 
the binding energies with an accuracy of up to three significant digits after 
the decimal point.

\begin{acknowledgments}
This work is supported by Russian Science Foundation (Grant No. RSF 23-22-00143).
\end{acknowledgments}

\end{document}